\documentclass[12pt]{article}
\usepackage{amssymb}
\usepackage{graphicx}
\author{ Jared Kaplan \footnote{jaredk@stanford.edu} \\ \textit { Department of Physics, Stanford University} }
\title{Extracting Data from Behind Horizons with the AdS/CFT Correspondence}
\begin{document} \maketitle

Recent work has shown that boundary correlators in
AdS-Schwarzschild can probe the geometry near the singularity.  In
this paper we aim to analyze the specific signatures of the
singularity, show how significant they can be, and uncover the
origins of these large effects in explicitly outside the horizon
descriptions.

We add perturbations to the metric localized near the singularity
and explore their effects on the boundary correlators.  Then we
use analyticity arguments to show how this information arises from
the Euclidean path integral of a free scalar field in the bulk.

\newpage

\subsection*{Introduction} In physics we are often able to understand a new object
by probing it with something we are already familiar with.  This
standard approach fails when applied to spacelike singularities
because our probes cannot get close to spacelike singularities and
live to tell about them - they become trapped behind horizons
longs before they approach the singularity itself. However, the
structure of a singularity can be probed by spacelike geodesics,
and in the AdS/CFT correspondence such geodesics approximate
opposite sided boundary correlators. These ideas helped motivate
previous work \cite{deta}, \cite{mald}, \cite{kos}, and
\cite{fhks} where it was shown that information about the
singularity could be extracted from these correlation functions.
Our purpose here is to show that these correlators really do
contain information about the geometry near the singularity, and
to give explicit examples demonstrating that it is `easy' to
access this data.  Most of these examples utilize the geodesic
approximation.

In \cite{kos} it was shown that there exists a dual description of
physics in $D=3$ AdS-Schwarzschild, depending on whether one
analytically continues the Euclidean metric with respect to
Schwarzschild or Kruskal time. In the Schwarzschild description,
all quantities can be determined with outside the horizon
calculations. Since we are extracting information about the
neighborhood of the singularity, it is surprising that we can see
large effects from outside the horizon computations. Thus our
final calculation will be an attempt to show how the large effects
arise from the full quantum path-integral.  The key will be
analyticity.

The layout of the paper is as follows. After a short review of the
AdS-Schwarzschild spacetimes, we use the WKB approximation to
explore the effects of a simple perturbation localized near the
black hole singularity in the $D=3$ case.  This is simply a small
change in the geometry near the singularity; it is not required to
satisfy Einstein's equation and we do not expect it to be a model
for anything physical.  Its purpose is to illustrate that we can
explore the geometry with boundary correlators by studying them as
a function of the boundary time. We find that the perturbation
produces a significant change in the correlators for relatively
small values of the boundary time, indicating that information
from behind the horizon is not difficult to extract, at least for
a meta-observer who can compute any gauge theory amplitude. Moving
on to a $D=5$ example, we make a very naive `stringy perturbation'
to the metric that resolves the singularity, and investigate the
consequences (this is not intended as a guess at what string
theory does, it is just a possibility). We find that a pole in the
opposite sided correlators resulting from geodesics that `bounce'
off the singularity in the unperturbed case splits into two poles
above and below the real axis.  This is a distinct, unambiguous
signature that we have removed the singularity.

Previous work has focused on the idea of behind the horizon data
made available from amplitudes computed outside the horizon, far
from the black hole. Thus as a concrete example, we calculate
approximately how many terms in a power series of the opposite
sided correlator as a function of boundary time one must compute
in order to see a large change resulting from a small perturbation
at the singularity.  We find that for a perturbation decaying as
$e^{-a r}$ it is only necessary to calculate $O(a)$ terms.
Finally, we set up the full quantum calculation of correlation
functions for a free scalar field in a perturbed Euclidean $D=3$
background.  We expand to first order in the perturbation and
derive a geodesic approximation, showing that in the Euclidean
case the quantum computation is consistent with the WKB
approximation.  This shows by virtue of the analyticity of
multi-point functions that the large effects of the perturbation
must be reproduced in path integral computations.

\subsection*{Spacetime Geometry} Here we will be considering the effects
of perturbations to the $D=3$ and $D=5$ AdS-Schwarzschild geometry
localized near the black hole singularity.  In both cases the
metric takes the form
\begin{equation}
ds^2 = -f(r)dt^2 + \frac{dr^2}{f(r)} + r^2 d \Omega^2
\end{equation}
Henceforth, we will ignore spherical part.  The function
\begin{equation}
f(r) = r^2-1-g(r)
\end{equation}
for the $D=3$ case and
\begin{equation}
 f(r) = r^2-\frac{1}{r^2}-g(r)
\end{equation}
for the $D=5$ case, with $g(r)$ a `small' perturbation function
for both cases.  It is essential that we always choose an analytic
$g(r)$ function to maintain equivalence with the boundary gauge
theory, whose correlation functions are analytic.  In these
backgrounds the geodesic equations take the form
\begin{eqnarray}
\left(\frac{dr}{ds}\right)^2 & = & E^2 + f(r) \\
\frac{dt}{ds} f(r) & = & E
\end{eqnarray}

The WKB approximation to the opposite-sided boundary two-point
function is given by $e^{-mL}$, where $L$ is the proper length of
the geodesic connecting the two points. Thus these geodesic
equations will be used extensively, beginning in the next section.

\begin{figure}
\begin{center}
\includegraphics[width=80mm, height=80mm]{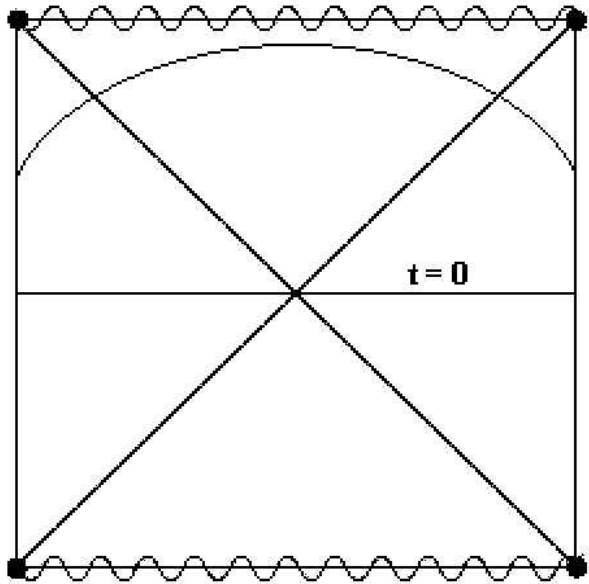}
\caption{\small{This figure shows the Penrose diagram for the
$D=3$ AdS-Schwarzschild spacetime with a symmetric spacelike
geodesic passing near the singularity.  The $D=5$ case has a
similar diagram, with one important difference - the singularities
must `bend outward', as shown in \protect\cite{fhks}.}}
\end{center}
\end{figure}

\subsection*{Some $D=3$ Examples}

\subsubsection*{Geodesics Far From the Singularity}

First we will give a very simple argument showing in general that
the effect of any $g(r)$ perturbation is appropriately small for
geodesics far from the singularity.  Integrating equation (4) we
find
\begin{equation}
L = \lim_{R_c \to \infty} \left[ 2 \int_{r_m}^{R_c}
\frac{dr}{\sqrt{E^2+f(r)}} - 2 \mathrm{Log}(R_c) \right]
\end{equation}
The factor of 2 comes from the two parts of a symmetric geodesic,
the $-$Log$(R_c)$ is a regulator, and $r_m$ is chosen to satisfy
$E^2+f(r_m) = 0$, so it is interpreted as the minimum of the $r$
coordinate along the geodesic curve. Since the perturbation $g(r)$
is localized near the singularity and decreases away from it, the
maximum of $g(r)$ is at $r_m$. But $r_m$ must itself be quite
large (of order 1 with our dimensionless setup) because the
geodesic was taken to be far from the singularity, so $g(r_m)$
must be extremely small. This gives a bound for $L$ (from now on
the regulator will be implicit)
\begin{equation}
2 \int_{r_m}^{\infty} \frac{dr}{\sqrt{E^2+r^2-1 - |g(r_m)|}} \leq
L \leq 2 \int_{r_m}^{\infty} \frac{dr}{\sqrt{E^2+r^2-1 +|g(r_m)|}}
\end{equation}
which gives
\begin{equation}
\mathrm{Log}\left(\frac{4}{-E^2+1 + |g(r_m)|}\right) \leq L \leq
\mathrm{Log}\left(\frac{4}{-E^2+1 - |g(r_m)|}\right)
\end{equation}
from which we see that $L$ doesn't change much for geodesics far
from the singularity, because such geodesics are characterized by
small $E$, and in that range there is no sensitivity to the tiny
$|g(r_m)|$ terms.  In other words, the change in $L$ is $ \sim
|g(r_m)|$.

\subsubsection*{Exponential Decay Models}

Now we will study a perturbation that decays exponentially as a
function of the $r$ coordinate.  More precisely, we take $g(r) =
\lambda^2 e^{-ar}$, where we will be taking $a \gg 1$.  In this
case, even at the black hole horizon, where $r=1$, the
perturbation will be exponentially small.

We are interested in correlation functions.  In the large mass
limit, we can use the WKB approximation, which is based on our
knowledge of
\begin{equation}
L = 2\int_{r_m}^{\infty} \frac{dr}{\sqrt{E^2+r^2-1-\lambda^2
e^{-ar}}}
\end{equation}
There is no hope of doing this integral explicitly.  It can be
accurately approximated if it is split into two integrals, but
this would
involve us in complex algebra that can be avoided by a more direct
approach. First note that in the standard, unperturbed case, as
the geodesic approaches the singularity, we have
\begin{equation}
L = 2\int_{r_m}^{\infty} \frac{dr}{\sqrt{r^2-r_m^2}} \to 2
\log(r_m)
\end{equation}
as $r_m \to 0$.  The exact solution in this case \cite{kos} gives
$L \approx t$ as $t \to \infty$, which is what we are seeing in
the equation above. However, with the perturbation turned on, we
get an integrand of the form $(Ar^2+ B r + C)^{-\frac{1}{2}}$ from
the expansion of the exponential at very small $r$, so both $L$
and $t$ are finite as $r_m \to 0$.

It is possible that we can recover a behavior similar to the
unperturbed case by continuing $r_m$ past $0$, and in fact, this
is exactly what happens.  The idea is this: we want to find an
$r_m$ that takes the integrals for $L$ and $t$ to $\infty$. Near
this $r_m$ the integrand must behave like $\frac{dx}{\sqrt{x^2}}$.
Thus we need to find a quadratic vanishing point of the function
under the square root, namely $E^2 + r^2 - 1 - \lambda^2 e^{-a
r}$.  This is easy to do, the result is that in the regime where
$a \lambda < \sqrt{\frac{2}{e}}$, we definitely have an $r_m$
value on the negative real r-axis. This is what we would like,
since it simply means the true singularity has `moved'. However,
outside of this regime we still have an $r_m$ that allows for
quadratic vanishing under the square root, but it will in general
be complex, giving a complex $E$ and generically complex $L$ and
$t$. Thus we will still find $L \approx c t$ for some $c$, but the
physical interpretation is less clear.

In the regime $a \lambda \ll 1$ we can approximate using
\begin{equation}
0 = E^2+r_m^2-1-\lambda^2 e^{-a r_m} \approx A \left(r_m +
\frac{B}{2A} \right)^2 + C - \frac{B^2}{4 A}
\end{equation}
by expanding the exponential to quadratic order and completing the
square. Now if we set $C = \frac{B^2}{4 A}$ by tuning $E$ we find
that $r_m = -\frac{B}{2 A}$. This allows us to approximate the
quadratic vanishing as
\begin{eqnarray}
L & = & 2 \int_{r_m}^{R_t} \frac{dr}{\sqrt{A \left(r +
\frac{B}{2A}
\right)^2 + C - \frac{B^2}{4 A}}} \nonumber \\
t & = & 2 \int_{r_m}^{R_t} \frac{E dr}{(1 - r^2 + \lambda^2 e^{-a
r})\sqrt{A \left(r + \frac{B}{2A} \right)^2 + C - \frac{B^2}{4
A}}}
\end{eqnarray}
where $R_t$ is some small value of $r$ where higher order terms
from the exponential become important.  Both of these integrals do
indeed diverge as we approach the $r_m$ noted above, and we can
roughly approximate $L \approx \left(1 + \frac{3 a^2 \lambda^4}{8}
\right) t$ for very large $t$ by setting the rest of the integrand
equal to its value at $r_m$. Thus we get a linear $L(t)$ for large
$t$ that is larger than in the unperturbed case.

The next example is $g(r) = \lambda^2 e^{-ar^2}$. We can use a
similar approximation scheme, but we will get different results -
$L \to \infty$ as $r_m \to 0$, so the analysis is simplified. This
is because when we expand the exponential we find
\begin{equation}
(1+\lambda^2 a)r_m^2 + (E^2-1-\lambda^2) = A r_m^2 + B \approx 0
\end{equation}
so the perturbation does not destroy the quadratic vanishing at
$r_m = 0$.  Again, it is possible to approximate the integral by
breaking it up, but a simpler analysis gives the large $t$
behavior.  The important point here is that this behavior comes
from the region very close to $r_m = 0$, so we can ignore the rest
of the integral.  We find
\begin{eqnarray}
t = 2 \int_{r_m}^{R_t} \frac{E dr}{\sqrt{Ar^2+B}(r^2-1-\lambda^2
e^{-a r^2})} & \approx & 2 \int_{r_m}^{R_t}
\frac{\sqrt{1+\lambda^2}
dr}{\sqrt{Ar^2+B}(1 + \lambda^2 - r^2(1+\lambda^2 a))} \nonumber \\
& \approx & \frac{L}{\sqrt{1+\lambda^2}}
\end{eqnarray}
So we see that for very small $r_m$ corresponding to very large
$t$, $L(t) \approx t \sqrt{1+\lambda^2}$, which is again larger
than in the unperturbed case.

A final question about the exponential decay models is at what
value of $t$ does the perturbed $L(t)$ differ significantly from
the original $L(t)$?  A quick way to get a rough estimate is to
use equation (5) to compare an infinitesimal perturbed geodesic
length element, $\delta s$, to an unperturbed length element,
$\delta s_0$, both at a point $r$:
\begin{equation}
\frac{\delta s}{\delta s_0} = \frac{r^2 - 1 - g(r)}{r^2 -1}
\end{equation}
Thus for a perturbation $g(r) = \lambda^2 e^{-a r^b}$, this
expression is significantly different from unity for $r <
a^{-1/b}$.  Here we interpret $r$ as $r_m$, the $r$ of closest
approach, so we see that the presence of $g(r)$ starts to make a
difference for $r_m^2 = 1-E^2 \approx a^{-2/b}$.  At this level of
approximation we can use the unperturbed expression for $t(E)$,
giving
\begin{equation}
t_s \approx \mathrm{Log} (4 a^{\frac{2}{b}} - 1)
\end{equation}
where $t_s$ is the value of $t$ where the perturbed and
unperturbed $L(t)$ functions begin to separate.  Even for very
large $a$ this is quite small, indicating that information about
the geometry near the singularity is not difficult to recover from
boundary correlators.

We can also approach the problem by estimating $\Delta L$, the
difference between the perturbed and unperturbed geodesic length,
and comparing it to the exact, unperturbed answer. To do this,
note that
\begin{equation}
\Delta L(E) = -2\int_{r_0}^{r_m} \frac{dr}{\sqrt{E^2 + r^2 - 1}}
 + 2\int_{r_m}^{\infty} \frac{dr}{\sqrt{E^2 + r^2 -
1-\lambda^2 e^{-ar^b}}} - \frac{dr}{\sqrt{E^2 + r^2 - 1}}
\end{equation}
where $r_0$ and $r_m$ are zeroes of $E^2 + f(r) = 0$ for the
unperturbed and perturbed cases, respectively.

We will bound the second term and then compute the first. A
similar result will be important in the $D=5$ case for different
purposes. The second term is smaller than
\begin{equation}
2\int_{r_m}^{\infty} \frac{dr}{\sqrt{E^2 + r^2 - 1-\lambda^2 e^{-a
r_m^b}}} - \frac{dr}{\sqrt{E^2 + r^2 - 1}}  =  2 \mathrm{Log}
\left(1 + \frac{\lambda^2 e^{-a r_m^b}}{r_m} \right)
\end{equation}
whereas the first term is just
\begin{equation}
-2 \mathrm{Log} \left(\frac{r_m + \lambda e^{-a r_m^b
/2}}{r_0}\right)
\end{equation} Since they contribute with opposite signs and the
absolute value of the first term is bigger, we see that (taking
absolute values)
\begin{equation}
2 \mathrm{Log} \left(\frac{r_m(r_m+\lambda e^{-a r_m^b /2})}{r_0
(r_m + \lambda^2 e^{-a r_m^b})}\right) < \Delta L(E)  <   2
\mathrm{Log} \left(\frac{r_m + \lambda e^{-a r_m^b
/2}}{r_0}\right)
\end{equation}
and both bounds are reasonable approximations for $\Delta L$
itself. Now we are interested in $\frac{|\Delta L|} {|L_0|}$, or
upon simplification
\begin{equation}
\frac{\log \left( \frac{r_m + \lambda e^{-a r_m^b
/2}}{r_0}\right)}{\log(2r_0)}
\end{equation}
Thus we see that for $r_m \sim a^{-\frac{1}{b}}$ and so $t \approx
\log(a)$, the perturbed $L$ differs from the unperturbed by a
sizable percentage, confirming our simpler method above.  It is
essential to remember that this method and the previous are only
estimates.  We have been calculating $L$ as a function of $r_m$,
not $t$, so we can only get a rough idea of when the perturbation
becomes important.

    Since both methods give an imprecise, but very encouraging
estimate for the $t$ value where the perturbation becomes
important, we will also give an upper bound above which the
perturbation must affect $L$ significantly.  We found the exact
asymptotic behavior for $L(t)$, and for both exponential models it
was $L \approx c t$ for large $t$, for some $c$, where $c=1$ in
the unperturbed case.  Once in this asymptotic regime, the
perturbation is certainly important.  This regime is characterized
by an $L$ dominated by the segment of the geodesic close to the
singularity. Noting that for $t=0$ we have $r_m = 1$ and $L = 2
\log(2)$, we see that if $L \gg 1$, $L$ must be dominated by
contributions from inside the horizon.  Since $L \varpropto t$,
for any $t \gg 1$ the perturbation makes a large effect, so in
particular, for $t \approx a$ the perturbation must be
significant.  Thus our previous analyses show that the
perturbation should be important for $t \approx \log(a)$, and this
calculation demonstrates that its effect will certainly have been
long established for $t \approx a$.

\subsubsection*{An Exactly Solvable Model}

The last perturbation to be considered for $D = 3$, $g(r) =
\frac{\lambda}{r^2}$, is really a very large perturbation near the
singularity, but it has the advantage that all integrals can be
done exactly.  Now we have
\begin{equation}
L = 2\int_{r_m}^{\infty} \frac{dr}{\sqrt{E^2+r^2 - 1 -
\frac{\lambda}{r^2}}}
\end{equation}
The integral is doable though nontrivial, defining $r_m$ as root
of the denominator of the integrand, we get the final result
\begin{equation}
L = \mathrm{Log}\left(\frac{4}{E^2-1+2r_m^2}\right)
\end{equation}
Note that as $r_m \to 0$, the $\frac{\lambda}{r^2}$ term implies
that $E \to \infty$, implying that $L \to -\infty$. This is
dramatically different from the unperturbed case because of the
minus sign, but this is not surprising since the effect is huge
near $r = 0$. The effect here is actually essentially the same as
in the standard $D=5$ model, which is obvious given the similar
behavior of this perturbation and the $D=5$ model near the
singularity. To see this more concretely, note that
\begin{equation}
t = 2\int_{r_m}^{\infty} \frac{E dr}{(r^2 -1 -
\frac{\lambda}{r^2})\sqrt{E^2+r^2 -1 - \frac{\lambda}{r^2}}}
\end{equation}
As noted above, the limit $r_m \to 0$ corresponds to $E \to
\infty$, so in this limit the $E$ dependent terms in the numerator
and denominator of the integrand cancel, and we have
\begin{equation}
t = 2\int_{0}^{\infty} \frac{dr}{r^2 -1 - \frac{\lambda}{r^2}}
\end{equation}
which is finite since there is no divergence from $r$ near $0$ or
$\infty$.  This gives a $t_c$ similar to that found by \cite{fhks}
in the $D=5$ case, where there is a pole in the WKB correlator
$e^{-m L}$ and the behavior of the $L(t)$ function changes
qualitatively. We can evaluate $t_c$ explicitly, simplification
gives
\begin{equation}
t_c = \frac{2 \pi}{\sqrt{2+8 \lambda}}\left(\sqrt{1+\sqrt{1+4
\lambda}} + \sqrt{\sqrt{1+ 4 \lambda} - 1} \right)
\end{equation}

\subsection*{Model Stringy Effects in $D = 5$}

Now we will consider a different sort of perturbation in the $D=5$
case.  The goal here is to examine how boundary correlators change
in a model where the curvature singularity at $r=0$ is `resolved',
so that curvature scalars are large and finite where they used to
be singular.  This could be considered as a model for stringy
effects that smear the singularity, but it is not intended as an
actual guess at what these effects are.  Thus we will take
\begin{equation}
f(r) = r^2 - \frac{1}{r^2+\epsilon^2}
\end{equation}
so that the $\epsilon = 0$ case corresponds to standard
AdS-Schwarzchild.  In the standard case there exists a special
value $t_c = \frac{\pi}{2}$ that corresponds to the geodesic
becoming nearly null, so that $L \to -\infty$ and the two point
correlator blows up.  We will find that this $t_c$ value is
`resolved' when $\epsilon \neq 0$, so that the pole in the
boundary correlator at $t_c$ splits into two poles, both off the
real $t$ axis.

It will be relevant in the following to note that when $t$ and $L$
are regarded as functions of $E$, the $t_c$ point corresponds to
$E \to \infty$. The integral representation of $t(E)$ for the
finite $\epsilon$ case can be parameterized as
\begin{equation}
t = -\int_{r_m^2}^{\infty} \frac{E dx \sqrt{x+\epsilon^2}}{(x -
\frac{1}{x+\epsilon^2}) \sqrt{x^3 + (E^2 + \epsilon^2) x^2 + (E^2
\epsilon^2 - 1)x}}
\end{equation}
Note that as $E \to \frac{1}{\epsilon}$, the coefficient of $x$
under the square root in the denominator vanishes.  Since in this
same limit we have $r_m = 0$, the integrand blows up as
$\frac{1}{x}$, so $t \to \infty$, which never occurs in the
unperturbed case. Thus for finite but very small $\epsilon$,
$t(E)$ increases steadily for a while, levels off near $t_c$, then
blows up to infinity for $E$ near $\frac{1}{\epsilon}$.  Note that
this divergence comes from the behavior of the integrand near
$x=0$, but the only difference between this integrand and the
integrand for the $L(E)$ function is the factor of $(x -
\frac{1}{x+\epsilon^2})$ in the denominator and the factor of $E$
in the numerator. These factors combine to give $L(t) \approx
\frac{t}{\epsilon}$ for $E$ near $\frac{1}{\epsilon}$, or $t \to
\infty$.

However, this leaves us with a mystery, because in the original
model $L(t) \approx 2 \mathrm{Log}(t-t_c)$ near $t_c$.  Where has
this $t_c$ singularity gone?  To answer this question we look at
\begin{equation}
L(E) = \int_{r_m^2}^{\infty} \frac{dx \sqrt{x +
\epsilon^2}}{\sqrt{x^3 + (E^2 + \epsilon^2) x^2 + (E^2 \epsilon^2
-1)x}}
\end{equation}
because we must have $L(t) \to -\infty$ at the new value(s) of
$t_c$.  For very large $E$ (much greater than
$\frac{1}{\epsilon}$) the integral expression for $L$ gets large,
because there is a large range of $x$ where the integrand is only
decaying like $x^{-1/2}$.

To see this explicitly, we can approximate $L(E)$ very roughly in
this regime as
\begin{equation}
L = \int_{-\epsilon^2}^{\epsilon^2} \frac{\epsilon dx}{\sqrt{E^2
x^2 + (E^2 \epsilon^2 -1)x}} + \int_{\epsilon^2}^{\infty}
\frac{dx}{\sqrt{x^2 + E^2 x + E^2 \epsilon^2 -1}}
\end{equation}
(the lower limit of integration is $-\epsilon^2$, not $0$, because
$-\epsilon^2$ is the largest root of $E^2 + f(r) = 0$) which gives
\begin{equation}
L = \frac{2 \epsilon}{E} \mathrm{Log} (2i + 2i E \epsilon) +
\frac{2 \epsilon^2}{E} \mathrm{Log}(2 E \epsilon + 2\sqrt{E^2
\epsilon^2 - 1}) - \mathrm{Log} \left( 2\epsilon^2 + E^2 +
2\sqrt{2E^2 \epsilon^2 -1} \right)
\end{equation}
The first two terms go to $0$ for large $E$, but the third term
goes to $-\infty$.  This divergence arises from the large $x$
behavior of the integrand, which explains why $t$ is finite for
large $E$ - its integrand includes an extra factor of
$\frac{1}{x}$.  Thus despite the singularity in $t$ at $E =
\frac{1}{\epsilon}$, it still makes sense to consider the limit $E
\to \infty$ in our search for the missing pole, because for very
large $E$, $L$ diverges while $t$ stays finite (close to the
original $t_c$ but slightly complex), imitating the $t_c$ behavior
of the unperturbed model. Calculating the full approximate value
of $t(E)$ is a bit lengthy, but the imaginary part of $t_c$ is
easier because it must come from the region where $x$ is negative.
Thus we can get it from
\begin{eqnarray}
\mathrm{Im}(t_c) & \approx & \mathrm{Im}\left(
\int_{-\epsilon^2}^{\epsilon^2} \frac{-E \epsilon
(x+\epsilon^2)}{\sqrt{E^2 x^2 + (E^2 \epsilon^2 - 1)x}} \right)
\nonumber
\\
& = & \pm \frac{i \pi \epsilon^3}{2}
\end{eqnarray}
where the $\pm$ comes from the choice of sign in the square roots
of negative quantities. This shows how a perturbation that
`resolves' the black hole singularity splits the Log$(t-t_c)$
pole. Without the perturbation we had $L(t) \approx 2
\mathrm{Log}(t-t_c)$ near $t_c$, so this must be modified at
finite $\epsilon$.  A good guess would be that $L(t) \approx
\mathrm{Log}((t-t_{c+})(t-t_{c-}))$, but this remains an estimate
because the integrals cannot be done algebraically.

\subsection*{Perturbations and Power Series}

In this section we study $f(r) = r^2 - \frac{1 + e^{-ar}}{r^2}$ in
the $D=5$ case.  In particular, we will analyze the power series
for $L(t)$ and show that the coefficients of this power series
differ from the coefficients of the unperturbed case by an amount
larger than the original coefficients after only $a$ terms. This
suggests that from a computational point of view our calculations
are surprisingly sensitive to the geometry near the singularity.
Though we choose this model for variety, our methods carry over
straightforwardly to the $D=3$ exponential models studied earlier.

First we note that for small $E$, $t(E) \approx E$, so we will
only worry about the $L(E)$ function - studying small $E$ is
equivalent to studying small $t$. This is an approximation,
because we are assuming that higher powers of $t$ in $E(t)$ do not
conspire to destroy the results we find. First consider the
unperturbed $L(E)$ function
\begin{equation}
L_0(E) = 2 \int_{r_0}^{\infty} \frac{dr}{\sqrt{E^2 + r^2 -
\frac{1}{r^2}}} = \mathrm{Log} \left( \frac{1}{\sqrt{1 +
\frac{E^4}{4}}}\right)
\end{equation}
where the denominator of the integrand is zero at $r_0$.  It will
be useful later to note that the coefficient of the $4n^{th}$ term
in the expansion of $r_0(E)$ is given approximately by $\frac{1}{n
4^n}$. Even more roughly, the coefficient of the $n^{th}$ power of
$E$ in the expansion of $L_0(E)$ is about $2^{-n/2}$.  Now define
\begin{equation}
\Delta L(E) = 2\int_{r_m}^{\infty} \frac{dr}{\sqrt{E^2 + r^2 -
\frac{1+e^{-ar}}{r^2}}} - 2\int_{r_0}^{\infty} \frac{dr}{\sqrt{E^2
+ r^2 - \frac{1}{r^2}}}
\end{equation}
where $r_m$ satisfies $E^2 + f(r_m) = 0$ in the perturbed case. We
will now examine the power series for $r_m$; this will be
important because later we will show that the coefficients of the
power series for $\Delta L(E)$ are essentially proportional to the
coefficients in the series for $r_m(E)$. We can write
\begin{equation}
r_m = 1 + x + \sum c_k E^k
\end{equation}
which we can plug into the defining equation to get
\begin{equation}
E^2(1 + x + \sum c_k E^k)^2 + (1 + x + \sum c_k E^k)^4 - 1 -
e^{-a-a x} e^{a\sum c_k E^k} = 0
\end{equation}
We can then use this equation to solve for the $c_k$ by equating
powers of $E$.  Assuming $x$ has been correctly computed the
coefficients of all the odd powers of $E$ will vanish (since the
equation defining $r_m$ only involves $E^2$).  We compute $c_2$ as
\begin{eqnarray}
E^2 + 4(1 + 3x)c_2 E^2 - e^{-a-a x} a c_2 E^2 = 0 \nonumber \\
\Rightarrow c_2 \approx \frac{1}{4}
\end{eqnarray}
Now we only need to keep the $c_2$ part when we expand the
exponential, because terms involving higher $c_n$ will have fewer
factors of $a$, which is a large number.  By expanding the
exponential we find
\begin{equation}
c_n \approx \frac{e^{-a} a^n}{n! 2^{n/2}} \approx \frac{e^{-a}
a^n}{n^n e^{-n} 2^{n/2}}
\end{equation}
What we are interested in is the ratio of these terms to the terms
in the expansion of $L_0$.  This ratio is $\frac{e^{-a} a^n}{n^n
e^{-n}}$, and it is maximized at the $a^{th}$ term, which is of
order $1$.

Now looking back to equation (34) and noting that $r_m > r_0$, we
can write
\begin{equation}
\Delta L(E) = 2\int_{r_0}^{r_m} \frac{-dr}{\sqrt{E^2 + r^2 -
\frac{1}{r^2}}}
 + 2\int_{r_m}^{\infty} \frac{dr}{\sqrt{E^2 + r^2 -
\frac{1+e^{-ar}}{r^2}}} - \frac{dr}{\sqrt{E^2 + r^2 -
\frac{1}{r^2}}}
\end{equation}
We will now show that the contribution of the second term to the
power series is exponentially suppressed as $e^{-a/2}$, and that
the first term contributes to the power series of $\Delta L(E)$
terms proportional to (or larger than) the $r_m$ series.  First we
bound the second term.  There is no theorem here that says that
our bound on the second term also bounds the power series in $E$,
but it should be extremely plausible in this case, because our
bound does not destroy any violent dependence on $E$.  In any
case, we can always just regard our bound as an approximation. It
is
\begin{eqnarray}
2^{nd} \mathrm{\ term} & < &  2\int_{r_m}^{\infty}
\frac{dr}{\sqrt{E^2 + r^2 - \frac{1+e^{-ar_m}}{r^2}}} -
\frac{dr}{\sqrt{E^2
+ r^2 - \frac{1}{r^2}}} \nonumber \\
& = &  \mathrm{Log} \left(1 + \frac{e^{-a r_m}} {A B} -
\frac{e^{-2a r_m}} {4A B^3} + ...\right)
\end{eqnarray}
where $B = r_m e^{-a r_m/2}$ and $A = E^2 + 2r_m^2 + 2B$.  Thus
every term is suppressed by $e^{-a r_m/2}$, which proves the claim
since $r_m$ is nearly $1$ for $t$ near $0$. Now we need only show
that the first term gives the much larger contribution claimed.
This is not very hard, because we can just evaluate the first term
directly, it is
\begin{eqnarray}
2\int_{r_0}^{r_m} \frac{-dr}{\sqrt{E^2 + r^2 - \frac{1}{r^2}}} & =
&  \mathrm{Log} \left(\frac{E^2 + 2r_m^2 + 2\sqrt{r_m^4 + E^2
r_m^2 -1}}{E^2 + 2 r_0^2}\right) \\ & = & \mathrm{Log} \left(1 +
\frac{2 r_0^2\sqrt{2r_m - 2 r_0}} {\sqrt{r_0 + r_0^5}} + \frac{4
r_0^2(r_m-r_0)}{1+ r_0^4} + ... \right) \nonumber
\end{eqnarray}
which when expanded gives a power series in $E$ with terms at
least as large as the $r_m$ series, because $r_0 \approx 1$. Thus
we have shown that after about $a$ terms the $\Delta L(E)$ power
series has coefficients that are large compared to the $L_0(E)$
coefficients.  This is noteworthy because the perturbation itself
is suppressed exponentially in $a$.

\subsection*{Path Integral Treatment}

All previous results have been obtained from the WKB
approximation, which applies in the large mass limit.  Thus it
would be interesting to see how large effects emerge from a small
perturbation in the full quantum mechanical treatment of the
problem. Furthermore, in \cite{kos} it was shown that there exists
a dual description of the $D=3$ AdS-Schwarzschild physics,
depending on whether Lorentzian correlators are obtained from the
Euclidean correlator through analytic continuation with respect to
the Schwarzschild time coordinate or Kruskal time. Thus it should
be possible to see the large effects in an explicitly outside the
horizon calculation. We will see how this is done, and the key
will be analytic continuation.

We wish to calculate the two point function for a massive free
scalar field in the perturbed S-AdS background.  Specifically, we
want to evaluate this two point function between field operators
on opposite sides of the Penrose diagram, with the black hole
between them, so that we can compare the effect of the
perturbations with the semi-classical treatment already given.  In
order to do this, we will compute in the Euclidean spacetime, and
then analytically continue in the $t$ coordinate.  The primary
object we will consider is the Euclidean path integral
\begin{equation}
I = \int D \Phi \ e^{- \int \frac{d^3x}{2} \left[
\frac{(\partial_t \Phi)^2}{r^2 -1 -g(r)} + (r^2 -1 -g(r))
(\partial_t \Phi)^2 + \frac{(\partial_{\phi} \Phi)^2}{r^2} + m^2
\Phi^2 \right]}
\end{equation}
Now for $g(r) = \lambda^2 e^{-a r^2}$ we can expand the argument
of the exponential to first order in $\lambda^2$, to get
\begin{equation}
I \approx \int D \Phi \ e^{-\int \frac{d^3x}{2} (\lambda^2
e^{-ar^2})\left[ \frac{1}{(r^2-1)^2} (\partial_t \Phi)^2  +
(\partial_r \Phi)^2 \right]} \ e^{-\int \frac{d^3x}{2} \left[
\frac{(\partial_t \Phi)^2}{r^2 -1} + (r^2 -1) (\partial_t \Phi)^2
+ \frac{(\partial_{\phi} \Phi)^2}{r^2} + m^2 \Phi^2 \right]}
\end{equation}
Note that the integrals are only evaluated outside the horizon,
since the behind the horizon region doesn't even exist in the
euclidean case.  Now we have approximated the first order effect
of the perturbation as an insertion of an operator smeared over
spacetime. Thus the two point function can be written as
\begin{eqnarray}
\langle \Phi(x_1) \Phi(x_2) \rangle & \approx & \left\langle
\Phi_0(x_1) \Phi_0(x_2) e^{-\int \frac{d^3x}{2} (\lambda^2
e^{-ar^2})\left[ \frac{1}{(r^2-1)^2} (\partial_t \Phi_0)^2  +
(\partial_r \Phi_0)^2 \right]} \right\rangle \\
& \approx & \left\langle \Phi_0(x_1) \Phi_0(x_2) \left( 1 - \int
d^3x \frac{\lambda^2 e^{-ar^2}}{2}\left[ \frac{1}{(r^2-1)^2}
(\partial_t \Phi_0)^2  + (\partial_r \Phi_0)^2 \right] \right)
\right\rangle \nonumber
\end{eqnarray}
The first term in the last line is obviously just the original,
unperturbed two point function, and the second term is the first
order effect of the perturbation.  This is equivalent to a
position dependent mass for the field, since it appears as the
integral of a function multiplied with the first derivative of the
field squared.  It can be calculated in the same way standard
Feynman diagrams are, meaning that we will `contract' each
exterior field with one of the squared differentiated fields. Just
as in a standard field field theory setting, we use $D(x, y)$, the
unperturbed propagator, to contract fields. This gives the second
term as
\begin{equation}
\Delta \langle \Phi(x_1) \Phi(x_2) \rangle \approx \int d^3x
\frac{\lambda^2 e^{-a r^2}}{2} \left[ -\frac{1}{(r^2-1)^2}
\partial_t D(x_1, x) \partial_t D(x_2, x) + \partial_r D(x_1,x) \partial_r D(x_2, x) \right]
\end{equation}
Now the $D=3$ AdS-Schwarzschild propagator is given by an infinite
sum, which arises from an infinite set of `images' of the AdS
propagator under the discrete identification of points that
creates the singularity.  To avoid this complication, we will be
forced to work in the $r_+ \to \infty$ limit, in which the
propagator takes the much simpler form
\begin{equation}
D = \left[\sqrt{\frac{r^2-r_+^2}{r_+^2}} \mathrm{cos}(r_+ \Delta
t) + \frac{r}{r_+} \mathrm{cosh}(r_+ \Delta \phi)\right]^{-2h_+}
\end{equation}
For definiteness and to avoid long expressions, we will only deal
with the $\partial_r D(x_1,x) \partial_r D(x_2, x)$ term (the
other term gives a result of the same order), and we will make a
few simplifying approximations along the way. This term is given
by
\begin{equation}
\int_0^{2 \pi} d \phi \int_C dt \int_{r_m}^{\infty} dr 2 h_+^2
\lambda^2 e^{-a r^2} r^{-4 h_+} \left(\frac{\cosh (\phi) -
\frac{r}{\sqrt{r^2-1}} \cosh (\Delta t_1)}{\left(\cosh (\phi) -
\frac{\sqrt{r^2-1}}{r} \cosh (\Delta t_1)\right)^{2h_+ +
1}}\right) (... \Delta t_2 ...)
\end{equation}
The contour $C$ is from $-T$ to $-2 \pi i - T$, but when we
analytically continue we will take $T \to \infty$ and $C$ will
move to avoid the singularities.  We can approximate $r_m$ as $1 +
\frac{\lambda^2 e^{-a}}{2}$, and $\Delta t_i = t - t_i$.

Our goal here is to make it clear how the geodesic approximation
gives a large contribution.  For this purpose it suffices to
derive an equivalent of the geodesic approximation from this
operator calculation. The $\phi$-angle dependence will not affect
any of the interesting dynamics, so for simplicity we set $\phi =
0$. The key step now is to write the integrand as an exponential,
giving
\begin{eqnarray}
\int_C dt \int_{r_m}^{\infty} dr \ 4 h_+^2 \lambda^2 e^{-m \left[2
\log ((r - \sqrt{r^2-1} \cosh (t-t_1))(r - \sqrt{r^2-1} \cosh
(t-t_2)))\right]} \nonumber \\ \times \ e^{-\left[a r^2 - \log
\left( \left( 1 - \frac{r}{\sqrt{r^2-1}}\cosh(t-t_1)
\right)\left(1 - \frac{r}{\sqrt{r^2-1}}\cosh(t-t_2) \right)
\right)\right]}
\end{eqnarray}
In the $m \to \infty$ limit the the second exponential is
irrelevant compared to the first.  Thus the integral should be
approximated by looking at the stationary point of this first
exponential. But the term being exponentiated is just
\begin{equation}
\log ((r - \sqrt{r^2-1} \cosh (t-t_1))(r - \sqrt{r^2-1} \cosh
(t-t_2))) \approx L_{1-x} + L_{x-2}
\end{equation}
where we define $L_{i-x}$ to be the geodesic length between the
boundary point $i$ and the integration point $x$.  One easy was of
seeing this is that
\begin{equation}
\frac{1}{(r-\sqrt{r^2-1} \cosh(\Delta t))^{2h_+}} \approx e^{-m
\log(r-\sqrt{r^2-1} \cosh(\Delta t))} \approx e^{-m L}
\end{equation}
Therefore taking the stationary point of this term is entirely
equivalent to taking the point $x$ to lie along the unperturbed
geodesic connecting the two boundary points. We conclude that in
the large $m$ limit we should integrate the perturbation along the
geodesic.  This gives the same result as the geodesic calculation
to first order in $\lambda$.  Since the perturbed two point
function is analytic in $\Delta t$, we see that the full quantum
calculation must give the same results as the geodesic
computation, including the large effect of the perturbation for
small $t$.

\subsection*{Conclusions}

We have seen in a variety of situations that data about the
geometry near the spacelike singularity of the AdS-Schwarzschild
spacetime is contained in boundary correlators.  In each case, the
recurring theme was that this information can be calculated more
easily than might be expected, and that it is sensitive to
specifics such as the function perturbing or `resolving' the
singularity.  This information is rather indirect, since we still
do not understand the spacelike singularities themselves. However,
it is possible that our method could be a clue to a more complete
description, and it is exciting to see explicit signatures of the
singularity.

\subsubsection*{Acknowledgements}
I would like to thank Lukasz Fidkowski, Veronika Hubeny, and
Matthew Kleban for discussions and insights, and the Kavli
Institute for Theoretical Physics at UCSB for their hospitality. I
was supported by the Summer Research College at Stanford
University, where I resided while this work was prepared. I
especially thank Stephen Shenker for his guidance and support for
this entire project.

\end{document}